\documentclass{article}

\input{tcilatex}

\begin{document}

\title{\textbf{Basic Quantum Theory and Measurement from the Viewpoint of Local
Quantum Physics }\\
{\small This is a condensed version of material prepared for and submitted
to proceedings of the symposium entitled ``New Insights in Quantum
Mechanics-Fundamentals, Experimental Results and Theoretical Directions''
Goslar, Germany, September 1-3, 1998~}}
\author{Bert Schroer \\
Institut f\"{u}r Theoretische Physik\\
FU-Berlin, Arnimallee 14, 14195 Berlin, Germany\\
presently: CBPF, Rua Dr. Xavier Sigaud, 22290-180 Rio de Janeiro, Brazil}
\date{March, 1999}
\maketitle

\begin{abstract}
Several aspects of the manifestation of the causality principle in LQP
(local quantum physics) are reviewed or presented. Particular emphasis is
given to those properties which are typical for LQP in the sense that they
do go beyond the structure of general quantum theory and even escape the
Lagrangian quantization methods of standard QFT. The most remarkable are
those relating causality to the modular Tomita-Takesaki theory, since they
bring in the basic concepts of antiparticles, charge superselections as well
as internal and external (geometric and hidden) symmetries.
\end{abstract}

\section{\protect\large LQP Principles and some Consequences}

If one thinks about the fundamental physical principles of this century
which have stood their grounds in the transition from classical into quantum
physics, relativistic causality as well as the closely related locality of
quantum operators (together with the localization of quantum states) will
certainly be the most prominent one.

This principle entered physics through Einstein's 1905 special relativity,
which in turn resulted from bringing the Galilei relativity principle of
classical mechanics into tune with Maxwell's theory of electromagnetism.
Therefore it incorporated Faraday's ``action at a neighborhood'' principle
which revolutionized 19$^{th}$ century physics.

The two different aspects of Einstein's special relativity, namely Poincar%
\'{e} covariance and the locally causal propagation of waves (in Minkowski
space) were kept together in the classical setting. In the adaptation of
relativity to \textbf{LQP} (local quantum physics\footnote{%
We use this terminology, whenever we want to emphasize that we relate the
principles of QFT not with necessarily with the standard text-book formalism
that is based on quantization through Lagrangian formalism.}) on the other
hand \cite{Haag}, it is appropriate to keep them at least initially apart in
the form of positive energy representations of the Poincar\'{e} group
(leading to Wigner's concept of particles) and Einstein causality of local
observables (leading to observable local fields and local generalized
``charges''). Here a synthesis is also possible, but it happens on a deeper
level than in the classical setting and results in LQP as a new physical
realm which is conceptually very different from both classical field theory
and general QT (quantum theory). The elaboration of some of these
differences, in particular as they may be relevant with respect to the
measurement process, constitutes one of the aims of these notes. For
material which already entered textbooks or review articles, we have
preferred to quote the latter. A more detailed account of the consequences
of causality in a much broader context can be found in \cite{Cau}\cite
{schroer}.

As a result of this added locality, LQP acquires a different framework than
the kind of general quantum theory setting \cite{Landsman} in which the
basics of quantum theory and measurement (including those ideas, which in
the fashionable language of the day, are referred to as ``quantum
computation'') are presented . Those concepts, which originate from the
quantum adaptation of Einstein causality, lead in the presence of
interactions to real particle creation (which artificially could be
incorporated into a multichannel version quantum theory of particles) and,
what has more importance within our presentation, to \textit{virtual
particle structure} (related to the phenomenon of vacuum polarization) which
has no counterpart in global general quantum theory as quantum mechanics and
cannot be incorporated into it at all. The latter remark preempts already
the greater significance of superselected charges and their fusion, as
opposed to particles and their quantum mechanical bound states. Thus the
hierarchy of particles in QM is replaced by the hierarchy of charges and
consequently we obtain ``nuclear democracy'' between particles. This is
closely related to an almost anthropological principle which LQP realizes in
a perfect way in laboratory particle physics: whenever energy-momentum and
(generalized) charge conservation allow for particle creation channels to be
opened, nature will maximally use this possibility. To be sure there are
theoretical models of LQP (integrable/factorizing models in d=1+1 spacetime
dimensions) which do not follow this dictum, but even in those cases at
least its theoretical ``virtual'' version is realized: a vector state
created by the application of an interacting field to the vacuum which has a
one-particle component, is inexorably accompanied by a ``polarization
cloud'' of particles/antiparticles (the hallmark of LQP). As already
emphasized the \textit{only exception} \textit{are free bosonic/fermionic
fields} and in a somewhat pointed (against history), but nevertheless
correct manner, one may say that this very exception is the reason why QM as
a nonrelativistic limit of LQP has a physical reality at all. More general
braid group statistics, as it can occur together with exotic spin in low
dimensional QFT, requires these polarization clouds already in the
``freest'' realization of anyons/plektons and they are not fading away in
the nonrelativistic limit because they are needed to uphold braid group
statistics in that limit. This is the reason why the attempts of Leinaas-
Myrheim, Wilscek and many others, which draw on the analogy with the
Aharanov-Bohm quantum mechanics may catch some aspects of plektons but miss
the spin-statistics connection which is their most important property (i.e.
their LQP characterization).

This aspect of virtuality, which at first sight seems to complicate life
since it activates the coupling between infinitely many degrees of
freedom/channels, is counterbalanced by some very desirable and useful
features: whereas general quantum theory needs an outside interpretative
support, LQP carries this already within itself. It was emphasized already
at the end of the 50$^{ies}$ (notably by Rudolf Haag \cite{Haag}), that e.g.
for a particle interpretation one does not need to resolve the distinction
between the various local observables which are localized in the same
space-time region (laboratory extension and time duration of measurement),
the knowledge of the space-time affiliation of a generic observable from a
region $\mathcal{O}$ is enough. The experimenter does not know more than the
geometric spacetime placement of his counters and their sensitivity; the
latter he usually has to determine by monitoring experiments. The basic
nature of locality in interpreting the particle aspect of a theory is
underlined by the fact that despite intense efforts nobody has succeeded to
construct a viable \textit{nonlocal} theory. Here ``viable'' is meant in the
sense of conceptual completeness, namely that a theory is required to 
\textit{contain its own physical interpretation} i.e. that one does not have
to invent or impose formulas from outside this theory.

Although physical reality may unfold itself like an onion or an infinite
Russian ``matrushka'' with infinitely many layers of ever more general
physical principles towards higher energies (smaller distances), it should
still continue to be possible to have a mathematically consistent theory in
each layer which is faithful to the principles valid in that layer. This has
been fully achieved for quantum mechanics, but this goal was not yet reached
in QFT. As a result of lack of nontrivial d=1+3 models or structural
arguments which could demonstrate that the physical locality and spectral
requirements allow for nontrivial solutions, the theory is still far from
conceptual maturity, despite its impressive perturbation successes in QED,
the Standard Model and in the area of Statistical Mechanics/Condensed Matter
physics.

Causality and locality are in a profound way related to the foundations of
quantum theory in the spirit of von Neumann, which brings me a little closer
to the topic of this symposium. In von Neumann's formulation, observables
are represented by selfadjoint operators and measurements are compatible if
the operators commute. The totality of all measurements which are relatively
compatible with a given set (i.e. noncommutativity within each set is
allowed) generate a subalgebra: the commutant $L^{\prime }$ of the given set
of operators $L$. In particular in LQP, a conceptual framework which was not
yet available to von Neumann, one is dealing with an isotonic ``net'' of
subalgebras (in most physically interesting cases von Neumann factors, i.e.
weakly closed operator algebras with a trivial center) $\mathcal{O}%
\rightarrow \mathcal{A}(\mathcal{O}).$ Therefore unlike quantum mechanics,
the spatial localization and the time duration of observables becomes an
integral part of the formalism. \textit{Causality gives an a-priori
information about the size} \textit{of spacetime }$\mathcal{O}$\textit{\
-affiliated operator (von Neumann) algebras:} 
\begin{equation}
\mathcal{A}(\mathcal{O})^{\prime }\supset \mathcal{A}(\mathcal{O}^{\prime })
\label{Einstein}
\end{equation}
in words: the commutant $\mathcal{A}(\mathcal{O})^{\prime }$ of the totality
of local observables $\mathcal{A}(\mathcal{O})$ localized in the spacetime
region $\mathcal{O}$ contains the observables localized in its spacelike
complement (disjoint) $\mathcal{O}^{\prime }.$ In fact in most of the cases
the equality sign will hold in which case one calls this strengthened
(maximal) form of causality ``Haag duality'' \cite{Haag}: 
\begin{equation}
\mathcal{A}(\mathcal{O})^{\prime }=\mathcal{A}(\mathcal{O}^{\prime })
\end{equation}
In words, the spacelike localized measurements are not only commensurable
with the given observables in $\mathcal{O}$, but every measurement which is
commensurable with all observables in $\mathcal{O},$ is necessarily
localized in the causal complement $\mathcal{O}^{\prime }.$ Here we extended
for algebraic convenience von Neumann's notion of observables to the whole
complex von Neumann algebra generated by hermitian operators localized in $%
\mathcal{O}.$ If one starts the theory from a net indexed by compact regions 
$\mathcal{O}$ as double cones, then algebras associated with unbounded
regions $\mathcal{O}^{\prime }$ are defined as the von Neumann algebra
generated by all $\mathcal{A}(\mathcal{O}_{1})$ if $\mathcal{O}_{1}$ ranges
over all net indices $\mathcal{O}_{1}\subset \mathcal{O}^{\prime }.$

Whereas the Einstein causality (\ref{Einstein}) allows a traditional
formulation in terms of pointlike fields $A(x)$ as 
\begin{equation}
\left[ A(x),A(y)\right] =0,\,\,\,\left( x-y\right) ^{2}<0,
\end{equation}
Haag duality can only be formulated in the algebraic net setting of LQP,
since it is not a property which can be expressed in terms of individual
operators. This aspect is shared by many other important properties and
results \cite{Haag}.

One can prove that Haag duality always holds after a suitable extension of
the net to the so-called dual net $\mathcal{A}(\mathcal{O})^{d}.$ The latter
may be defined independent of locality in terms of relative commutation
properties as 
\begin{equation}
\mathcal{A}(\mathcal{O})^{d}:=\bigcap_{\mathcal{O}_{1},\mathcal{O}%
_{1}^{\prime }\subset \mathcal{O}}\mathcal{A}(\mathcal{O}_{1})^{\prime }
\end{equation}
The relative commutance with respect to the observables is called
(algebraic) ``localizability''. These considerations show that causality,
locality and localization in LQP have a natural and deep relation to the
notion of compatibility of measurements. In addition there are subtle
modifications with respect to the basic quantum structure with possible
changes of environmental and other aspects of quantum measuring. The
fundamental reason for all such modifications in the interpretation of LQP
versus QM is the different structure of local algebras: the vacuum is not a
pure state with respect to any algebra which is equal to or contained in an $%
\mathcal{A}(\mathcal{O})$ with $\mathcal{O}^{\prime }$ nonempty, and the
sharply localized algebras $\mathcal{A}(\mathcal{O})$ themselves do not
admit pure states at all\footnote{%
In order to find local algebras which are anywhere near quantum mechanical
algebras and admit pure states and tensor products with entanglement similar
to the inside/outside quantization box situation in Schr\"{o}dinger theory,
one has to allow for a ``fuzzy'' transition ``collar'' between a double cone
and its causal disjoint outside, in more precise terms one has to consider a
so-called split inclusion \cite{Haag}.}! They possess an algebraic structure
which has not been taken into account in the present day presentation of
quantum basics including quantum computation. Since these fine points can
only be appreciated with some more preparation, I will postpone their
presentation.

If the vacuum net (i.e. the vacuum representation of the observable net) is
Haag dual, then all associated ``charged'' nets share this property, unless
the charges are nonabelian (in which case the deviation from Haag duality is
measured by the Jones index of the above inclusion, or in physical terms the
statistics- or quantum-dimension \cite{S-W}). If on the other hand even the
vacuum representation of the observable net violates Haag duality, then this
indicates spontaneous symmetry breaking \cite{Roberts} i.e. not all internal
symmetry algebraic automorphisms are spatially implementable. As already
mentioned, in that case one can always maximize the algebra without
destroying causality and without changing the Hilbert space, such that Haag
duality is restored. This turns out to be related to the descend to the
unbroken part of the symmetry which allows (since it is a subgroup) more
invariants i.e. more observables.

Since QM and what is usually referred to as the basics of quantum theory do
not know these concepts at all, I am presenting in some sense a contrasting
program to the (global) QT orientation of this symposium. But often one only
penetrates the foundations of a framework more profoundly, if one looks at a
contrasting structure even if the difference is (presently) not measurable.
For an analogy we may refer to the Hawking effect which has attracted ever
increasing attention as a matter of principle, even though there is hardly
any experimental chance.

In connection with this main theme of this symposium, it is interesting to
ask if LQP could add something to our understanding of classical versus
quantum reality (the ERP, Bell issue) or the measurement process i.e.
production of ``Schr\"{o}dinger cat states'' and observation of their
subsequent decoherence. For the first issue I refer to \cite{Summers}. Apart
from some speculative remarks \cite{Landsman}, there exists no investigation
of the measurement process which takes into consideration the characteristic
properties of the local algebras in LQP. I tend to believe that, whereas
most of the present ideas on coherent states of Schr\"{o}dinger cats and
their transition to von Neumann mixtures will remain or at least not suffer
measurable quantitative modifications, LQP could be expected to lead to
significant conceptual changes. Certainly it will add a universal aspect to
the issue of decoherence through environments. Contrary to QM where the
environment is introduced by extending the system, localized systems in LQP
are always open subsystems for which the ``causal disjoint'' defines a kind
of universal environment which is build into its formalism.

Another structurally significant deviation which was already alluded to
results from the fact that the vacuum becomes a thermal state with respect
to the local algebras $\mathcal{A}(\mathcal{O}).$ There are two different
mechanisms to generate thermal states: the standard coupling with a heat
bath and the thermal aspect through restriction or localization and the
creation of horizons \cite{Sew}\cite{Verch}. The latter is in one class with
the Hawking-Unruh mechanism; the difference being that in the localization
situation the horizon is not classical i.e. is not defined in terms of a
differential geometric Killing generator of a symmetry transformation of the
metric.

The fact that algebras of the type $\mathcal{A}(\mathcal{O})$ have no pure
states is related to the different behavior of the pair inside/outside with
respect to factorization: whereas in QM the boxed system factorizes with the
system outside the box, the total algebra $B(H)$ in LQP is generated by $%
A(O) $ and its commutant $B(H)=A(O)\vee A(O)^{\prime },$ but \textit{it is
not the tensor product} of the two factor algebras $\mathcal{A}(\mathcal{O})$
and $\mathcal{A}(\mathcal{O})^{\prime }=\mathcal{A}(\mathcal{O}^{\prime }).$
In order to get back to a tensor product situation and be able to apply the
concepts of entanglement and entropy, one has to do a sophisticated split
which is only possible if one allows for a ``collar'' (see later) between $%
\mathcal{O}$ and $\mathcal{O}^{\prime }$ \cite{Haag}.

Since the thermal aspects of localization are analogous to black holes%
\footnote{%
The analogy is especially tight for the wedge localization since the
boundary of wedges define bifurcated classical ``Killing horizons'' (Unruh),
whereas the boundary of e.g. a double cone in a massive theory defines a
``quantum horizon''. This concept has a cood meaning with respect to the
nongeometrically acting modular group associated with the latter situation,
and it has no classical analogon (it is in fact a ``hidden symmetry'').},
there is no chance to directly measure such tiny effects. However in
conceptual problems, e.g. the question if and how not only classical
relativistic field theory, but also QFT excludes superluminal velocities,
these subtle differences play a crucial role. Because of an unusual property
of the vacuum in QFT (the later mentioned Reeh-Schlieder property), the
exclusion of superluminal velocities requires more conceptual and
mathematical understanding than in the classical case. Imposing the usual
algebraic structure of QM (i.e. assuming tacitly that the local observables
allow pure states) onto the local photon observables will lead to
nonsensical results. Most sensational theoretical observations on causality
violations which entered the press and in one case even Phys. Rev. Letters,
suffer from incorrect tacit assumptions (if they are not already caused by a
misunderstanding of the classical theory). We urge the reader to look at the
fascinating reference \cite{Yng-Buch} and the conceptually wrong preceding
article.

Historically the first conceptually clear definition of localization of
relativistic wave function was given by Newton and Wigner \cite{New Wig} who
adapted Born's x-space probability interpretation to the Wigner relativistic
particle theory. Apparently the result that there is no exact satisfactory
relativistic localization (but only one sufficient for all practical
purposes) disappointed Wigner so much, that he became distrustful of the
usefulness of QFT in particle physics altogether (private communication by
R. Haag). Whereas we know that this distrust was unjustified, we should at
the same time acknowledge his stubborn insistence in the importance of the
locality concept which he thought of as an indispensable requirement in
addition the positive energy property and irreducibility of the Wigner
representations. Without explanation we state that modular localization of
state vectors is different from the Born probability interpretation. Rather
subspaces of modular localized wave functions preempt the existence of
causally localized observables already on the level of the Hilbert space of
relativistic wave functons and have no counterpart at all in N-particle
quantum mechanics. As will be explained later, modular localization may
serve as a starting point for the construction of interacting
nonperturbative LQP's \cite{S-W}\footnote{%
In fact the good modular localization properties are guarantied in finite
component positive energy representations, with the Wigner infinite
component ``continuous spin'' representations being the only exception.. In
this infinite component finite energy representation it is not possible to
come from the wedge localization down to the spacelike cone localization
which is the coarsest localization which one needs for a particle
interpretation.}. It is worthwhile to emphasize that sharper localization of
local algebras in LQP is not defined in terms of support properties of
classical smearing functions but via the rather unusual formation of
intersection of localized algebras; although in some cases as CCR- or
CAR-algebras (or more generally Wightman fields) the algebraic formulation (%
\ref{Einstein}) can be reduced to this more classical concept.

Since the modular structure is related to the so-called KMS property \cite
{Haag}, it is not surprising that the modular localization has thermal
aspects. In fact as mentioned before, there are two manifestations of
thermality, the standard heat bath thermal behavior which is described by
Gibbs formula or, after having performed the thermodynamic limit, by the KMS
condition, and thermality caused by localization either with classical
bifurcated Killing-horizons as in black holes \cite{Sew}\cite{Verch} curved
spacetime and (Rindler, Unruh, Bisognano-Wichmann) wedge regions, or in a
purely quantum manner as the boundary of the Minkowski space double cones.
In the latter case the KMS state has no natural limiting description in
terms of a Gibbs formula (which only applies to type $I$ and $II$, but not
to type $III$ von Neumann algebras), a fact which is also related to the
boundedness from below of the Hamiltonian, whereas the e.g. Lorentz boost
(the modular operator of the wedge) does not share this property. In \cite
{Jaekel} the reader also finds an discussion of localization and cluster
properties in a heat bath thermal state. Although in these notes we will not
enter these interesting thermal aspects, it should be emphasized that
thermality (similar to the concept of virtual particle clouds) is an
inexorable aspect of localization in LQP and does not need the Hawking type
of Killing vector horizons. The close relation of particle and thermal
physics (KMS thermal property$\simeq $crossing symmetry of S-matrix and
formfactors \cite{S-W}) is a generic property of LQP and should not be
counted as a characteristic success of string theory.

Already in the very early development of algebraic QFT \cite{Haag-Sch} the
nature of the local von Neumann algebras became an interesting issue.
Although it was fairly easy (and expected) to see that i.e. wedge- or double
cone- localized algebras are von Neumann factors (in analogy to the tensor
product factorization of standard QT under formation of subsystems, it took
the ingenuity of Araki to realize that these factors were of type $III$
(more precisely hyperfinite type $III_{1},$ as we know nowadays thanks to
the profound contributions of Connes and Haagerup), at that time still an
exotic mathematical structure. Hyperfiniteness was expected from a physical
point of view, since approximatability as limits of finite systems (matrix
algebras) harmonizes very well with the idea of thermodynamic+scaling limits
of lattice approximations. A surprise was the type $III_{1}$ nature which,as
already mentioned, implies the absence of pure states (in fact all
projectors are Murray von Neumann equivalent to 1) on such algebras; this
property in some way anticipated the thermal aspect (Hawking-Unruh) of
localization. Overlooking this fact (which makes local algebras
significantly different from QM), it is easy to make conceptual mistakes
which could e.g. suggest an apparent breakdown of causal propagation \cite
{Yng-Buch} as already mentioned before. If one simply grafts concepts of QM
onto the causality structure of LQP (e.g. quantum mechanical tunnelling,
structure of states) without deriving them in LQP , one runs the risk of
wrong conclusions about e.g. the possibility of superluminal velocities.

A very interesting question is: what is the influence of the always present
causally disjoint environment on the measurement process, given the fact
that in the modern treatment the coupling to the environment and the
associated decoherence relaxation are very important. Only certain aspects
of classical versus quantum reality, as expressed in terms of Bell's
inequalities, have been discussed in the causal context of LQP \cite{Summers}%
. In the following we will sketch some more properties which set apart QM
from LQP and whose conceptual impacts on decoherence of Schr\"{o}dinger
cats, entanglement etc. still is in need of understanding.

Let me mention two more structural properties, intimately linked to
causality, which distinguish LQP rather sharply from QM. One is the
Reeh-Schlieder property: 
\begin{eqnarray}
\overline{\mathcal{P}(\mathcal{O})\Omega } &=&H,\,\,i.e.\,\,cyclicity\,\,of%
\,\,\Omega   \label{cyc} \\
A\in \mathcal{P}(\mathcal{O}),\,\,A\Omega  &=&0\Longrightarrow
A=0\,\,i.e.\,\,\Omega \,\,separating  \nonumber
\end{eqnarray}
which either holds for the polynomial algebras of fields or for operator
algebras $\mathcal{A}(\mathcal{O}).$ The first property, namely the
denseness of states created from the vacuum by operators from arbitrarily
small localization regions (a state describing a particle behind the moon%
\footnote{%
This weird aspect should not be held against QFT but rather be taken as
indicating that localization by a piece of hardware in a laboratory is also
limited by an arbitrary large but finite energy, i.e. is a ``phase space
localization'' (see subsequent discussion). In QM one obtains genuine
localized subspaces without energy limitations.} and an antiparticle on the
earth can be approximated inside a laboratory of arbitrary small size and
duration) is totally unexpected from the global viewpoint of general QT and
has even attracted the interest of philosophers of natural sciences. If the
naive interpretation of cyclicity/separability in the Reeh-Schlieder theorem
leaves us with a feeling of science fiction, the way out is to ask: which
among the dense set of localized states can be really produced with a
controllable expenditure (of energy)? In QM to ask this question is not
necessary since, as already mentioned, the localization at a given time via
support properties of wave functions leads to a tensor product factorization
of inside/outside so that the ground state factorizes and the application of
the inside observables never leads to a dense set in the whole space. It
turns out that most of the very important physical and geometrical
informations are encoded into features of dense domains and in fact the
aforementioned modular theory is explaining such relations. For the case at
hand, the reconciliation of the Reeh-Schlieder theorem with common sense has
led to the discovery of the physical relevance of \textit{localization with
respect to phase space in LQP}, i.e. the understanding of the size of
degrees of freedom in the set: 
\begin{eqnarray}
P_{E}\mathcal{A}(\mathcal{O})\Omega \,\,is\;compact &&\,\,\, \\
\,\,e^{-\beta \mathbf{H}}\mathcal{A}(\mathcal{O})\Omega
\;is\,\,nuclear,\,\,\,\,\mathbf{H} &=&\int EdP_{E}\,\,  \nonumber
\end{eqnarray}
The first property was introduces way back by Haag and Swieca \cite{Haag},
whereas the second statement (and similar nuclearity statements involving
modular operators of local regions instead of the global Hamiltonian) which
is more informative and easier to use, is a later result of Buchholz and
Wichmann \cite{Haag}. It should be emphasized that the LQP degrees of
freedom counting of Haag-Swieca, which gives an infinite (but still nuclear)
number of localized states is different from the finiteness in QM, a fact
often overlooked in present day's string theoretic degree of freedom
counting. The difference to the case of QM decreases if one uses instead of
a strict energy cutoff a Gibbs damping factor $\ e^{-\beta H}.$ In this case
the map $\mathcal{A}(\mathcal{O})\rightarrow e^{-\beta H}\mathcal{A}(%
\mathcal{O})\Omega $ is ``nuclear'' if the degrees of freedom are not too
much accumulative in order to prevent the existence of a maximal (Hagedorn)
temperature. The nuclearity assures that a QFT, which was given in terms of
its vacuum representation, also exists in a thermal state. An associated
nuclearity index turns out to be the counterpart of the quantum mechanical
Gibbs partition function \cite{Haag} and behaves in an entirely analogous
way.

The peculiarities of the above Haag-Swieca degrees of freedom counting are
very much related to one of the oldest ``exotic'' and at the same time
characteristic aspects of QFT: vacuum polarization. As discovered by
Heisenberg, the partial charge: 
\begin{equation}
Q_{V}=\int_{V}j_{0}(x)d^{3}x=\infty 
\end{equation}
diverges as a result of uncontrolled vacuum fluctuations near the boundary.
For the free field current it is easy to see that a better definition
involving test functions, which takes into account the fact that the current
is a 4-dim distribution and has no restriction to equal times, leads to a
finite expression. The algebraic counterpart is the already mentioned so
called ``split property'' namely \cite{Haag} that if one leaves between say
the double cone (``relativistic box'') observable algebra $\mathcal{A}(%
\mathcal{O})\,$and its causal disjoint $\mathcal{A}(\mathcal{O}^{\prime })$
a ``collar'' region, then it is possible to construct in a canonical way a
type $I$ tensor factor $\mathcal{N}$ which extends into the collar and one
obtains inside/outside factorization if one leaves out the collar region (a
fuzzy box). This is then the algebraic analog of Heisenberg's smoothening of
the boundary to control vacuum fluctuations. It is this ``split inclusion''
which allows to bring back some of the familiar structure of QM, since type
I factors allow for pure states, tensor product factorization, entanglement
and all the other properties at the heart of quantum theory and the
measurement process. Although there is no time to explain this, let us
nevertheless mention that the most adequate formalism for LQP which
substitutes quantization and is most characteristic of LQP in
contradistinction to QT, is the formalism of modular localization related to
the Tomita modular theory of von Neumann algebras. The interaction enters
through wedge algebras, thus giving wedges a similar fundamental role as
they already had in the Unruh illustration of the thermal aspects of the
Hawking effect. Modular localization also leads to a vast enlargement of the
symmetry concepts in QFT \cite{Hidden}\cite{Buch} beyond those geometric
symmetries which enter the theory through quantized Noether currents.

If by these remarks I have created the impression that local quantum physics
is one of the conceptually most fertile and spiritually (not historically)
young areas of future basic research with relevance to the basics of
measurement and quantum computation, I have accomplished the purpose of
these notes. Indeed I know of no other framework which brings together such
seemingly different ideas as Spin \& Statistics, TCP and crossing symmetry
of particle physics on the one hand together with thermal and entropical
aspects of (modular) localization \& black hole physics on the other hand.

\end{document}